# Mach's Principle in the Acoustic World

Ion Simaciu[1a], Gheorghe Dumitrescu[2], Zoltan Borsos[1]

[1] Petroleum-Gas University of Ploieşti, Ploieşti 100680, Romania

[2] High School Toma N. Socolescu, Ploieşti, Romania

## Abstract

*The aim of this paper is to investigate the coupled oscillations of multiple bubbles within a cluster. The interaction between a bubble and the other bubbles in a cluster produces an additional mass. For a fixed number of bubbles ( $N \gg 1$ ) and uniformly distributed, in case of a certain value of the bubble density, we deduce the relations analogous to the Eddington relation (between the cluster radius and the bubble radius) and the Sciama relation (between the cluster radius and the gravitoacoustic radius) according to Mach's Principle.*

Keyword: the acoustic world, bubbles cluster, Mach's Principle, Large numbers hypothesis

PACS: 43.25.+y, 43.35.+d, 47.55.dd

## 1. Introduction

Previous study of phenomena induced by acoustic waves in a fluid revealed an analogy between the Acoustic World and the Electromagnetic World [1].

Acoustic waves transport energy and mass [2] and generate wave packets and bubbles [3]. Consequently, this yields a rise of the fluid's inhomogeneities and, subsequently, a shift of the refractive index of the fluid. The effect of fluid non-homogeneity in the presence of a wave packet and a bubble is the deviation of an acoustic plane wave [1, 4, 5]. The oscillating bubbles interact through the secondary Bjerknes forces [6]. These forces are analogous to the electrostatic forces for the interaction generated by the scattering of sound waves [7] and are analogous to gravitational forces for the interaction generated by the absorption of sound waves [8]. The secondary Bjerknes forces are responsible for generating the clusters of bubbles [9].

As we will show in the following, some interesting effects can occur when the bubbles within a cluster oscillate in a coupled manner [10-12]. One of the consequences is a growth in virtual mass of the oscillating bubble when *N* identical bubbles oscillate. This additional mass depends on the parameters of the bubble and of the liquid (bubble radius $R_0$ and the volume density of the fluid $\rho$ ) and on the parameters of the cluster (cluster radius $R_c$ and density of bubbles $n_c$ or the average distance between the bubbles $d_c = n_c^{-1/3}$ ).

When the number of the bubbles in the cluster increases, this additional mass becomes significantly larger than that of the virtual mass of the bubble performing a radial oscillation.

In the second section of this paper we will derive the additional mass, mentioned above, when the bubbles within a cluster perform coupled oscillations.

In the third section we will point out that the additional mass can induce certain effects to electrostatic kind of acoustic interactions. We also highlight the existence of analogous relations with the Eddington relation [14, 15] and the Sciama relation [16, 17] in accordance with Mach's Principle.

[a] isimaciu@yahoo.com

Finally, some conclusions are listed in the last section.

## 2. Dynamics of coupled bubbles within a cluster

### 2.1. The issue of the interaction of N bubbles

The dynamics of interacting bubbles within a cluster can be described with the extended Rayleigh-Plesset equations [10-12]:

$$R_i \ddot{R}_i + \frac{3}{2}\left(1 - \frac{\dot{R}}{3u}\right)\dot{R}_i^2 = \frac{1}{\rho}\left[\left(p_0^\infty + \frac{2\sigma}{R_i} - p_\upsilon\right)\left(\frac{R_{0i}}{R_i}\right)^{3\kappa} - \frac{2\sigma}{R_i} + p_\upsilon - \frac{4\mu \dot{R}_i}{R_i} - \left(p_0^\infty - p_a \sin \omega t\right)\right] - \sum_{j \neq i} \frac{R_j^2 \ddot{R}_j + 2 R_j \dot{R}_j^2}{r_{ij}}. \quad (1)$$

where

$$p_{is} = \rho \sum_{j \neq i} \frac{R_j^2 \ddot{R}_j + 2 R_j \dot{R}_j^2}{r_{ij}} \cong \rho \sum_{j \neq i} \frac{R_j^2 \ddot{R}_j}{r_{ij}} \quad (2)$$

the the pressure resulting from the radiation scattered and absorbed by the other bubbles in the clusters, $r_{ij}$ are the lengths between the centers of the bubbles $i$ and $j$, $p_\upsilon$ is the inner pressure of the bubble, $p_0^\infty = p_0$ is the pressure of the liquid away from the bubbles, $\sigma$ is the surface tension coefficient of the liquid, $\mu$ is the dynamic viscosity coefficient, $R_{0i}$ is the equilibrium radius of the $i$ bubble.

When oscillations have small amplitude $R_i = R_{0i}\left[1 + x(t)\right] = R_{0i}(1 + x)$, therefore

$$\dot{R}_i = R_{0i} \dot{x}, \quad \ddot{R}_i = R_{0i} \ddot{x}. \quad (3)$$

When omitting the pressure $p_{is}$ in the equation (1), i.e. the linearized case, then the equation has the form

$$\ddot{x} + 2\beta_{ir} \dot{x} + \omega_{0ir}^2 x = -\frac{A}{\rho R_{0i}^2} \cos \omega t, \quad (4)$$

where natural angular frequency is

$$\omega_{0ir} = \left[3\gamma\left(\frac{p_0}{\rho R_{0i}^2} + 2\frac{\sigma}{\rho R_{0i}^3}\right) - 2\frac{\sigma}{\rho R_{0i}^3} + \frac{\omega^4 R_{0i}^2}{u^2\left(1 + \omega^2 R_{0i}^2/u^2\right)}\right]^{1/2} \cong \frac{1}{R_{0i}}\left(\frac{p_{eff}}{\rho}\right)^{1/2}, \quad (5)$$

the whole damping is

$$\beta_{ir} = \beta_{\upsilon ir} + \beta_{thir} + \beta_{acir}, \quad (6)$$

where radial viscous component $\beta_{\upsilon ir}$, radial thermal component $\beta_{thir}$ and radial acoustic component $\beta_{acir}$ are

$$\beta_{\upsilon r} = 2\frac{\mu}{\rho R_{0i}^2}, \beta_{thr} = 2\frac{\mu_{th}}{\rho R_{0i}^2}, \beta_{acr} = \frac{\omega^2 R_{0i}}{2u\left(1 + \omega^2 R_{0i}^2/u^2\right)} \cong \frac{\omega^2 R_{0i}}{2u}. \quad (7)$$

When replacing Eq. (3) in Eq. (2), this yield

$$p_{is} = \rho \sum_{j \neq i}^{N} \frac{R_{0j}^3 \left[(1+x_j)^2 \ddot{x}_j + 2(1+x_j)\dot{x}_j^2\right]}{r_{ij}} \cong \rho \sum_{j \neq i}^{N} \frac{R_{0j}^3 (1+x_j)^2 \ddot{x}_j}{r_{ij}} \cong \rho \sum_{j \neq i}^{N} \frac{R_{0j}^3 \ddot{x}_j}{r_{ij}}. \tag{8}$$

Assuming that bubbles are identical $R_{01} = R_{02} = ... = R_{0i} = R_0$, then Eq. (8) becomes

$$p_{is} = \rho R_0^3 \sum_{j \neq i} \frac{\ddot{x}_j}{r_{ij}}. \tag{9}$$

Equation (4) to which we add the pressure of the scattered radiation (9) becomes a system of equations [10, 11]

$$\ddot{x}_i + 2\beta_{ir}\dot{x}_i + \omega_{0ir}^2 x_i = -\frac{A}{\rho R_{0i}^2}\cos\omega t - \frac{1}{R_{0i}^2}\sum_{j \neq i}^{N} \frac{R_{0j}^3 \ddot{x}_j}{r_{ij}}. \tag{10}$$

For identical bubbles, the systems of equations (10) becomes

$$\ddot{x}_i + 2\beta_{ir}\dot{x}_i + \omega_{0ir}^2 x_i = -\frac{A}{\rho R_{0i}^2}\cos\omega t - \left(\sum_{j \neq i}^{N} \frac{R_{0i}}{r_{ij}}\right)\ddot{x}_i \tag{11a}$$

or

$$\left(1 + \sum_{j \neq i} \frac{R_0}{r_{ij}}\right)\ddot{x}_i + 2\beta_r \dot{x}_i + \omega_{0r}^2 x_i = -\frac{A}{\rho R_0^2}\cos\omega t. \tag{11b}$$

When comparing Eqs. (4) and (11b) it is observed that the radiation pressure scattered by the other cluster bubbles induces an additional inertial term $N_c = \sum_{j \neq i}^{N}(R_0/r_{ij}) > 0$, i.e. the inertial mass of each oscillating bubble increases from $m = 4\pi R_0^3 \rho$ at

$$m_N = 4\pi R_0^3 \rho \left(1 + \sum_{j \neq i}^{N} \frac{R_0}{r_{ij}}\right) = m + mN_c, \quad N_c = \sum_{j \neq i}^{N} \frac{R_0}{r_{ij}}. \tag{12}$$

Thus, the radial oscillation's inertia for each bubble is also determined by the interaction with other bubbles

$$m_N = m + mN_c = m + m_c \cong m_c, m_c = mN_c, N_c \gg 1. \tag{13}$$

We note that induced mass, $mN_c$, is proportional to the virtual mass of the bubble. According to the arguments from the section 3.2., we call the number $N_c$ the acoustic Mach's number.

It follows that it is not possible to induce an extra/aditional mass if the virtual mass of the bubble is zero. This condition is also found for the phenomenon of induction of mass in the Electromagnetic World [18].

The increase in the mechanical inertia of each bubble within the cluster leads to changing the magnitudes of the other parameters of the bubble (see section 2.2).

## 2.2. Oscillations of the bubbles within a cluster

In order to deduce the oscillations' parameters of identical bubbles whitin a cluster, we proceed to divide Eq. (11b) against the coefficient of acceleration $\ddot{x}_i$ in order to obtain the equation of the forced oscillator

$$\ddot{x}_i + 2\frac{\beta_r}{\left(1 + \sum_{j \neq i} \frac{R_0}{r_{ij}}\right)}\dot{x}_i + \frac{\omega_{0r}^2}{\left(1 + \sum_{j \neq i} \frac{R_0}{r_{ij}}\right)}x_i = -\frac{A}{\rho R_0^2 \left(1 + \sum_{j \neq i} \frac{R_0}{r_{ij}}\right)}\cos\omega t \tag{14}$$

or

$$\ddot{x}_N + 2\beta_{Nr}\dot{x}_N + \omega_{0Nr}^2 x_N = -\frac{A_N}{\rho R_0^2}\cos\omega t, \tag{15}$$

where we used the following parameters, according also to (12),

$$\omega_{0Nr}^2 = \frac{\omega_{0r}^2}{1+N_c} < \omega_{0r}^2, \quad \beta_{Nr} = \frac{\beta_r}{\left(1+\sum_{j\neq i}\frac{R_0}{r_{ij}}\right)} = \frac{\beta_r}{1+N_c} < \beta_r, \quad A_N = \frac{A}{1+N_c} < A. \tag{16}$$

One can see from the above relations that, according to the assumption made in section 2.1, the increase of the virtual mass induces a decrease in its own natural angular frequency $\omega_{0Nr}^2 < \omega_{0r}^2$, a decrease of the damping coefficient $\beta_{Nr} < \beta_r$ as well as a decrease in the amplitude of oscillations $A_N < A$ [13].

Assuming the solution of Eq. (15) is of the form $x_N = a_N \cos(\omega t + \varphi_N)$, it results, according to [6]

$$a_N = \frac{A_N}{\rho R_0^2\left[(\omega^2-\omega_{0Nr}^2)^2 + 4\beta_{Nr}^2\omega^2\right]^{1/2}}, \quad \varphi_N = \arctan\frac{2\beta_{Nr}\omega}{(\omega^2-\omega_{0Nr}^2)}. \tag{17}$$

Replacing Eqs. (16) in Eq. (17) yields

$$a_N = \frac{A}{\rho R_0^2(1+N_c)\left[\left(\omega^2-\frac{\omega_{0r}^2}{1+N_c}\right)^2 + \frac{4\beta_r^2\omega^2}{(1+N_c)^2}\right]^{1/2}}, \quad \varphi_N = \arctan\frac{2\beta_r\omega}{(1+N_c)\left(\omega^2-\frac{\omega_{0r}^2}{1+N_c}\right)} \tag{18a}$$

or

$$a_N = \frac{A}{\rho R_0^2\left[(\omega^2(1+N_c)-\omega_{0r}^2)^2 + 4\beta_r^2\omega^2\right]^{1/2}}, \quad \varphi_N = \arctan\frac{2\beta_r\omega}{\omega^2(1+N_c)-\omega_{0r}^2}. \tag{18b}$$

At the resonance of velocity, $\omega = \omega_{0Nr} = \omega_{0r}/\sqrt{1+N_c} < \omega_{0r}$, the amplitude and the phase (18b) become

$$a_{Nr} = \frac{A}{2\rho R_0^2 \beta_{rr}\omega_{0Nr}}, \quad \varphi_{Nr} = \arctan\infty = \frac{\pi}{2}, \tag{19}$$

where $\beta_{rr} = \beta_r(\omega = \omega_{0Nr})$ and

$$\beta_{rr} = \beta_{\upsilon r} + \beta_{thrr} + \beta_{acrr} = \frac{2\mu}{\rho R_0^2} + \frac{2\mu_{th}(\omega=\omega_{0Nr})}{\rho R_0^2} + \frac{\omega_{0Nr}^2 R_0}{2u}. \tag{20}$$

Special case $\beta_{acrr} \gg \beta_{\upsilon r} + \beta_{thrr}$ yields $\beta_{rr} \cong \beta_{acrr} = \omega_{0Nr}^2 R_0/(2u)$, therefore the amplitude displayed in Eq. (19) becomes

$$a_{Nrac} = \frac{Au}{\rho R_0^2 \omega_{0Nr}^3} = \frac{Au(1+N_c)^{3/2}}{\rho R_0^3 \omega_{0r}^3} = \frac{Au(1+N_c)^{3/2}}{p_{eff}^{3/2}} = \frac{A(1+N_c)^{3/2}}{p_{eff}}\left(\frac{\rho u^2}{p_{eff}}\right)^{1/2} \tag{21a}$$

or, for $N_c \gg 1$,

$$a_{Nrac} = \frac{A(1+N_c)^{3/2}}{p_{eff}} \left(\frac{\rho u^2}{p_{eff}}\right)^{1/2} \cong \frac{A N_c^{3/2}}{p_{eff}} \left(\frac{\rho u^2}{p_{eff}}\right)^{1/2}. \quad (21b)$$

When resonance occurs, one can see from Eq. (21b) that amplitude $a_{Nrac}$ depends on the parameters of the bubble and of the cluster through $N_c$.

## 2.3. Cluster with uniform distribution of bubbles

We assume a spherical cluster with radius $R_c \gg R_0$ having a large number of bubbles, $N \gg 1$,

$$N = \frac{4\pi R_c^3}{3} n_c, \quad (22)$$

with uniformly distributed bubbles and the number density $n_c \cong 1/d_c^3$ (here $d_c$ is the average distance between the bubbles), results

$$N = \frac{4\pi R_c^3}{3 d_c^3}. \quad (23)$$

The sum $N_c = \sum_{j \neq i}^{N} (R_0/r_{ij})$ expressed in Eq. (12) can be approached through integral calculation

$$N_c \cong R_0 \int_{d_c}^{R_c} \frac{n_c dV}{r} = R_0 \int_{d_c}^{R_c} 4\pi n_c r dr = 2\pi R_0 n_c (R_c^2 - d_c^2) = \frac{2\pi R_0 (R_c^2 - d_c^2)}{d_c^3} = \frac{2\pi R_0 R_c^2}{d_c^3} - \frac{2\pi R_0}{d_c} \cong \frac{2\pi R_0 R_c^2}{d_c^3}. \quad (24)$$

In the following, we will assume that the test bubble is placed in the center of the cluster and it interacts with $dN = n_c dV$ bubbles contained in a spherical layer having thickness $dr$ and surface $4\pi r^2$.

For $R_c > d_c > R_0$, $N_c$ is less than $N$

$$N_c \cong \frac{2\pi R_0 (R_c^2 - d_c^2)}{d_c^3} = \frac{4\pi R_c^3}{3 d_c^3} \frac{3}{2}\left(\frac{R_0}{R_c} - \frac{R_0 d_c^2}{R_c^3}\right) = N \frac{3}{2}\left(\frac{R_0}{R_c} - \frac{R_0 d_c^2}{R_c^3}\right) \cong N \frac{3 R_0}{2 R_c} < N. \quad (25)$$

When considering the distance from the bubble to the center of the cluster, we notice a force acting on each bubble, which is directed towards the center of the cluster (Eq. 11 from [12]). This force is similar to the electrostatic force acting on a uniformly distributed charge or to a gravitational force acting on a uniformly distributed mass at a distance less than that of the radius of the distributed charge or mass ($r_0 < r_{clust} = R_c$). This issue will be a topic for a further paper.

## 3. Acoustic interactions in a cluster

### 3.1. Interaction of electrostatic type

According to [6], the acoustic force for two identical bubbles, $R_{01} = R_{02} = R_0$, is

$$F_B(r,\varphi) \cong -\frac{2\pi \rho \omega^2 R_0^6}{r^2} a^2, \cos\varphi = 1. \quad (26)$$

Replacing Eqs. (18) in Eq. (26), with $a = a_N$, yields

$$F_{BN}(r) \cong \frac{-2\pi\omega^2 R_0^2 A^2}{\rho r^2 \left[\left(\omega^2(1+N_c) - \omega_0^2\right)^2 + 4\beta_r^2 \omega^2\right]} \cong \frac{-2\pi\omega^2 A^2 R_0^2}{\rho r^2 \left[\left(\omega^2 N_c - \omega_0^2\right)^2 + 4\beta_r^2 \omega^2\right]}, \; N_c \gg 1. \quad (27)$$

When resonance occurs, $\omega^2 = \omega_{0Nr}^2 \cong \omega_{0r}^2 / N_c$, then

$$F_{BNr}(r) \cong \frac{-\pi A^2 R_0^2}{2\rho r^2 \beta_r^2(\omega_{0N})} = \frac{-\pi A^2 R_0^2}{2\rho r^2 \left[\dfrac{2\mu}{\rho R_0^2} + \dfrac{2\mu_{th}(\omega = \omega_{0Nr})}{\rho R_0^2} + \dfrac{\omega_{0Nr}^2 R_0}{2u}\right]^2} \cong$$

$$\frac{-2\pi A^2 u^2}{\rho r^2 \omega_{0Nr}^4} = \frac{-2\pi A^2 u^2 N_c^2}{\rho r^2 \omega_{0r}^4}. \quad (28)$$

Adopting the natural angular frequency from Eq. (5) then Eq. (28) changes into

$$F_{BNr}(r) \cong \frac{-2\pi A^2 \rho u^2 R_0^4 N_c^2}{r^2 p_{eff}^2} = N_c^2 F_{Br}(r) \gg F_{Br}(r), \quad (29)$$

i.e. a much higher force is obtained than that acting between two bubbles that are not found in a cluster.

## 3.2. The Mach's Principle in Acoustic World

The analysis of the mechanical inertia induction phenomenon by coupling the oscillations of the bubbles in a cluster (section 2.2) leads to the hypothesis that this phenomenon is analogous to the one modeled by Sciama [16] for the mechanical inertia in the Universe (Electromagnetic World), according to Mach's Principle.

The additional mass term, according to Eqs. (12) and (13), is a term according to the Mach's Principle [16, 17]: *mechanical inertia, as a property of a body, is also determined by the interaction with the other bodies in the Universe*. In this case, the interaction between the bubbles that oscillate in phase is of electroacoustic nature. This interaction propagates with the velocity of acoustic waves such that $R_c = u\tau_c$.

For a continuous distribution of bubbles into a cluster, according to Eq. (25), the rate of inertia increase is the acoustic Mach's number

$$N_c = N\frac{3R_0}{2R_c} \cong \frac{NR_0}{R_c} = \frac{NR_0}{u\tau_c}, \; R_c \gg R_0. \quad (30)$$

This relationship, for $N_c = N^{1/2}$, is analogous to the relationship between the radius of the universe and the electromagnetic radius of the electron obtained by Eddington [14, 15], $R_U = N_U^{1/2} R_e$,

$$R_c N_c = NR_0 \rightarrow R_c N^{1/2} = NR_0 \rightarrow R_c = N^{1/2} R_0, \quad (31)$$

that is, there is a connection between the radius of the cluster and the radius of the bubble (electroacoustic radius).

The Eddington relation together with the relation obtained by Sciama (Eq. 7 from paper [16]) for gravitational interaction of particles in the Universe (Electromagnetic World)

$$Gn_U m_U \tau^2 \cong \frac{Gm_U}{c^2} \frac{N_U}{R_U^3}(c\tau)^2 = \frac{N_U R_g}{R_U} \cong 1, \; R_U = c\tau, \quad (32)$$

involves a relationship between the electromagnetic radius of the electron and the gravitational radius of the proton (the gravitational radius of the proton is $R_g = Gm_U/c^2$ )

$$R_e = N_U^{1/2} R_g. \qquad (33)$$

By analogy, we hypothesize that in the Acoustic Universe, there is also a relationship between the radius of the bubble (electroacoustic radius) and the gravitoacoustic radius of bubble

$$R_0 = N^{1/2} R_{ag}. \qquad (34)$$

Substituting this relation in the eqution (31) we obtain a Sciama relation in the Acoustic World

$$R_{cg} = N R_{ag}. \qquad (35)$$

We noted this radius with the index $g$, because this is the gravitoacoustic radius of the cluster. It follows that the condition $N_c = N^{1/2}$ involves the compression of the cluster to its gravitoacustic radius.

For an Acoustic Universe consisting of identical bubbles (analogous to the toy universe model proposed by Ibison [19]), the two radii correspond to the same kind of particle. For this reason, we will use relation (34) in a future paper to identify the gravitational mass and the gravitational constant for the gravitoacoustic interaction.

## 4. Conclusions

The study of the coupling oscillations of the bubbles, contained within a cluster, reveals the effect of increase the inertia of any bubble from the cluster. This increase is due to the scattered acoustic radiation by the other bubbles of the cluster.

For a large number of identical bubbles uniformly distributed, the induced mass, $mN_c$, causes the decrease of the natural angular frequency ( $\omega_{0Nr}^2 < \omega_{0r}^2$ ) and the damping coefficient ( $\beta_{Nr} < \beta_r$ ) and, implicitly, the modification of the expression of the electroacoustic force between two bubbles inside the cluster.

If the cluster is compressed so that the acoustic Mach's number is $N_c = N^{1/2}$, the surprising result of relations analogous to the Eddington and Sciama relations is obtained.

The results of this paper support the hypothesis of the analogy between the physical-mathematical models of the Acoustic World and the Electromagnetic World. These results suggest that in the Electromagnetic World the internal oscillations of the fundamental particles are coupled through the electromagnetic interaction.